\documentclass{aa}
\usepackage{graphics} 
\usepackage{psfig}

\begin{document}    
\thesaurus{06     
(08.09.2 \object{RZ Leonis } 
08.14.2;  
08.06.3;  
08.05.3;  
08.02.3)  
}

\title{The dwarf nova RZ Leonis : photometric period, ``anti-humps" and 
normal alpha disk$^{*}$}
   \thanks{Based on observations obtained at Las Campanas Observatory and ESO La 
Silla Observatory (ESO Proposal 54.E-0812)}
   \author{R.E.\ Mennickent\inst{1}\fnmsep\thanks{On leave in Harvard-Smithsonian 
Center for Astrophysics, 60 Garden St, MA 02138, Cambridge, USA}            
   \and        
   C.\ Sterken\inst{2}
   \and
   W.\ Gieren\inst{1}
\and
E.\ Unda\inst{1}} 
   \offprints{R.E.\ Mennickent}    
\institute{Universidad de Concepci\'on,
         Departamento de F\'{\i}sica,  Casilla 4009, Concepci\'on, Chile.\\          
         \and
         University of Brussels (VUB), Pleinlaan 2, 1050 Brussels, Belgium.\\
             }
   \date{Received ... ; accepted ...}    \titlerunning{The dwarf nova RZ Leonis 
}  
\authorrunning{Mennickent et al.}    \maketitle 

\begin{abstract}

We present results of differential photometry of the dwarf nova \object{RZ Leonis 
} 
spanning a 11-year baseline. The most striking feature
of the light curve is a non-coherent periodic hump of variable amplitude. 
A seasonal time series analysis
yields a photometric period of 0\fd0756(12). 
In addition, low amplitude fluctuations of the mean magnitude in time scale of 
months 
are observed. We find that the hump's amplitude is anti-correlated with 
the star's mean magnitude and becomes ``negative" (i.e.\ an absorption feature or 
``anti-hump") 
when the system is very faint. Secondary humps and ``anti-humps" are also 
observed.
The transition from ``anti-humps" to fully developed humps occurs on a time scale
of 70 days. We interprete the observations as a rapid response of the accretion 
disk to 
the increase of mass transfer rate.  In this case we deduce a viscosity parameter 
$\alpha$ $\sim$ 0.08,
i.e.\ much larger than often claimed for WZ~Sge-like stars. We note that the 
secondary
star in \object{RZ Leo} is close to a  main-sequence red dwarf and not a
brown-dwarf like star as suggested for other long cycle-length SU UMa stars like
\object{WZ Sge} and \object{V592 Her}.  Our results indicate that 
large amplitude and long cycle length  dwarf novae might not necessarily 
correspond to objects in the same evolutive stage.

\end{abstract}

\section{Introduction: about \object{RZ Leonis }}

Dwarf novae are interacting binary stars in which a Roche-lobe filling 
main-sequence
secondary  looses mass through the $L_{1}$ point. The transferred mass falls along 
a ballistic
trajectory towards the heavier white dwarf primary,  forming an accretion disk. 
The disk undergoes semi-periodic collapse during which matter is accreted by the 
compact primary. The result is a release of gravitational energy which is observed 
as a system brightening. This is called a dwarf nova outburst. 
Dwarf novae, a subclass of cataclysmic variable stars (CVs),
have been reviewed by Warner (1995a).

\object{RZ Leonis } is a long cycle-length large-amplitude dwarf nova 
with only 7 outbursts recorded since 1918 (e.g\ Vanmuster \& Howell 1996) and with
an estimated distance from earth between 174 and 246 pc (Sproats et al.\ 1996).
 

Humps in the light curve of \object{RZ Leo} repeating with a 0\fd0708(3) period 
were observed by Howell \& Szkody (1988).
They concluded that this dwarf nova is a candidate for SU UMa star probably seen 
under a large inclination.
This assumption is supported by the finding of broad double emission-lines in the 
optical spectrum
(Cristiani et al.\,1985,  Szkody \& Howell 1991). Orbital humps are observed in  
some high-inclination
dwarf novae (e.g.\ Szkody 1992), they probably reflect the pass of the disk-stream 
interacting region
(often named hot spot or bright spot) along the observer's line of sight.

The study of the hot spot variability of RZ Leo is potentially useful to constrain 
models 
of gas dynamics in close binary systems. In the classical view, 
the hot spot is formed during the shock interaction of matter in the gaseous 
stream flowing
from $L_{1}$ (the inner Lagrangian point) with the outer boundary of the accretion 
disk.
This picture was consistent with photometric observations of dwarf novae during 
many years. 
However, this view conflicts with recents observations indicating anomalous hot 
spots in many
systems. For example, in many cases, Doppler tomography does not show the effect 
of a 
hot spot at all, or indicates that the hot spot is not in the place where we would 
expect a collision
between the gaseous stream and the outer boundary of the disk (e.g.\ Wolf et al.\ 
1998). 
To explain these findings, the possibility of gas stream overflow has been worked 
out. In this view the hot spot is
formed behind the white dwarf by the ballistic
impact of a deflected stream passing  over the white dwarf
(e.g.\ Armitage \& Livio 1998, Hessmann 1999). 
However, this scenario has not yet been confirmed by observations. Furthermore, 
recent three-dimensional numerical simulations
indicate the {\it absence} of a shock between the stream and the disk. The interaction 
between the stream and the
{\it common envelope} of the system forms and extended shock wave along the edge 
of the stream, 
whose observational properties are roughly equivalent to those of a hot spot in 
the disk (Bisikalo et al.\ 1998). 

This paper is aimed to confirm the reported photometric period and to establish a
long-term hump ephemeris. We also expect to detect systematic luminosity trends 
and get insights
about the hump variability and hot spot nature.  Interestingly, we find some 
phenomena
conflicting,  in many ways, with the classical scenario of the hot spot forming 
region.

\begin{figure}
\centerline{\hbox{
\psfig{figure=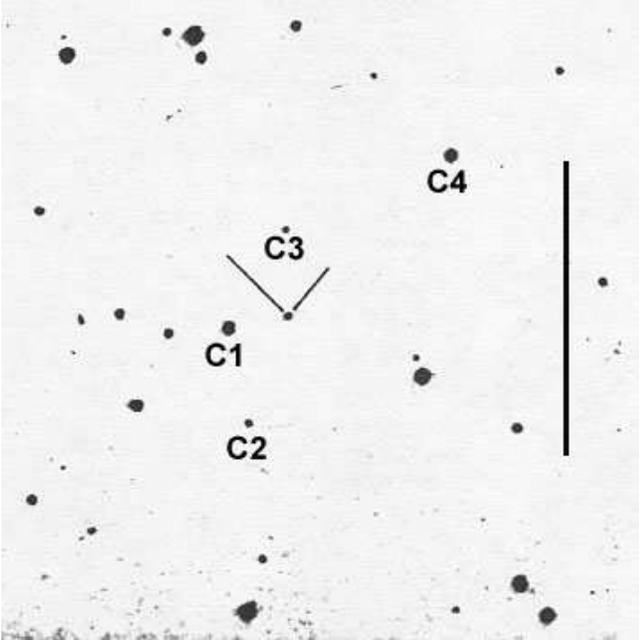,height=8.5cm,width=8.5cm,angle=0}
}}
\caption[]{The finding chart for RZ Leo. North is up  and East left. 
Comparison and check stars are labeled. The vertical bar shows a 5-arcminute 
distance. Adapted
from Vogt \& Bateson (1982).
}
\end{figure}

\begin{table*}
\caption[]{Journal of observations. N is the number of science frames per night. 
HJD is the heliocentric julian day at the start of every sequence,
referred to the zero point 244\,0000. 
Comparison ($C$) and check ($CH$) stars are labeled accordingly to Fig.\,1.
The variances of the $C-CH$ differences and $V-C$ light curve and the mean 
magnitude of RZ Leo
are also given. Note the decrease of variability associated to the faint state
of Jan-Feb 1998. Setup $A$ refers to the  1.0\,m LCO telescope and $B$ to the  
Dutch 0.92\,m ESO telescope.}
\begin{center}
\begin{tabular}{|c|r|r|c|c|c|c|c|c|} \hline
\multicolumn{1}{c}{Date (UT)} &
\multicolumn{1}{c}{HJD }&
\multicolumn{1}{c}{N} &
\multicolumn{1}{c}{setup}&
\multicolumn{1}{c}{$C$} &
\multicolumn{1}{c}{$CH$}  &
\multicolumn{1}{c}{$\sigma_{C-CH}$} &
\multicolumn{1}{c}{$\sigma_{V-C}$} &
\multicolumn{1}{c}{$\overline{V}$} \\
\hline \hline
18/03/91& 8333.5981  &58 &$A$ &C1   &C3  &0\fm06 &0\fm14   &18.42\\
19/03/91& 8334.6296  &44 &$A$ &C1   &C3  &0\fm08 &0\fm14   &18.45\\
20/03/91& 8335.7690  &19 &$A$ &C2   &C3  &0\fm04 &0\fm15   &18.47\\
11/03/95& 9787.7102  &46 &$A$ &C4   &C3  &0\fm03 &0\fm13   &18.80\\
07/01/98 &10820.7488 &25 &$B$ &C1   &C2  &0\fm02 &0\fm04   &19.08\\
11/01/98 &10824.7719 &27 &$B$ &C1   &C2  &0\fm02 &0\fm05   &19.13\\
06/02/98 &10850.7540 &30 &$B$ &C1   &C2  &0\fm01 &0\fm03   &19.06\\
07/02/98 &10851.7520 &30 &$B$ &C1   &C2  &0\fm02 &0\fm04   &18.94\\
18/03/98 &10890.6939 &36 &$B$ &C1   &C2  &0\fm02 &0\fm09   &18.71\\
19/03/98 &10891.6219 &22 &$B$ &C1   &C2  &0\fm02 &0\fm08   &18.67\\
22/01/99 &11200.7394 &26 &$B$ &C2   &C3  &0\fm02 &0\fm07   &18.71\\
23/01/99 &11201.7722 &25 &$B$ &C2   &C3  &0\fm02 &0\fm08   &18.63\\
\hline \hline
\end{tabular}
\end{center}
\end{table*}

\section{The observations and data reduction}

CCD images were obtained at six observing runs 
during 1991--1999 at Las Campanas Observatory (LCO) 
and ESO La Silla Observatory, Chile. Exposure times were between
250 and 300 s. Details of the observations are given in Table 1. 

All science images were corrected for bias and were flat-fielded 
using standard IRAF\footnote{IRAF is distributed by the National Optical 
Astronomy Observatories, which are operated by the Association of Universities
for Research in Astronomy, Inc., under cooperative agreement with the National 
Science Foundation} routines.  Instrumental
magnitudes were calculated with the {\it phot} aperture photometry package,
which is adequate due to the \object{RZ Leo}'s
uncrowded field.  The optimum aperture radius defined by Howell (1992)
was used. This radius matches the $HWHM$ of the point spread function, 
minimizing the noise contribution due to sky pixels and readout noise. 

In this paper we are interested in differential photometry.
This technique, reviewed by Howell (1992), involves the determination of
time series $V - C$ and $C - CH$, among instrumental magnitudes 
of variable ($V$), comparison ($C$) and
check ($CH$) stars in the same CCD field.
A finding chart of \object{RZ Leo} showing the check and 
comparison stars is shown in Fig.\,1.

The photometric error of $V - C$ was derived from the 
standard deviation of the $C - CH$ differences. In general, 
the intrinsic variance (not due to variability but 
to noise) associated to each differential light curve $\sigma_{V-C}$ 
and $\sigma_{C-CH}$ are related by a scale factor $\Gamma$ depending on the 
relative brightness of the sources (Howell \& Szkody\ 1988, Eq.\,13). 
This factor is of order of unity 
if the three sources are of similar brightness or if
the variable is of similar brightness to the
check star and the comparison is brighter. These criteria
are completely fulfilled in our observations.

Table 1 shows mean $V$ magnitudes 
along with the comparison and check stars used every night. 
The star labeled $C1$ in Fig.\,1 (for which $V$ = 14.201
is given by Misselt 1996), was used to shift
the differences to an non-differential magnitude scale.
On the other hand, $UBV$ magnitudes 
taken at HJD 244\,8333.5981, 244\,8333.6044 and 244\,8333.6131 
were properly calibrated with photometric standard stars, yielding $V$ = 18\fm56
$\pm$ 0\fm04, $B-V$ = 0\fm17 $\pm$ 0\fm07 and $U-B$ = -1\fm02 $\pm$ 0\fm08.

\section{Results}

\begin{figure}
\centerline{\hbox{
\psfig{figure=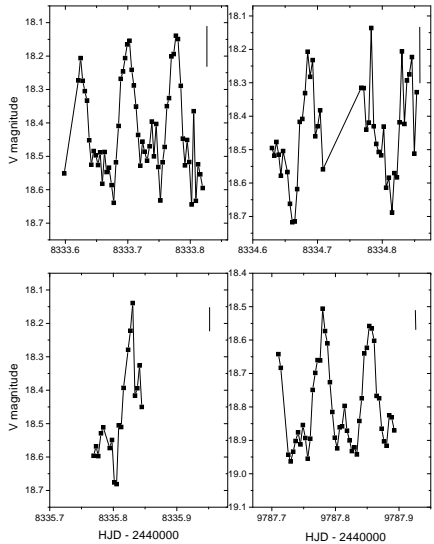,height=3.5in,angle=0}
}}
\caption[]{Differential magnitudes 
on March 1991 and 1995. Note the high amplitude humps. A typical error is shown
by  the bar in the upper right corner of every panel.}
\end{figure}

\begin{figure}
\centerline{\hbox{
\psfig{figure=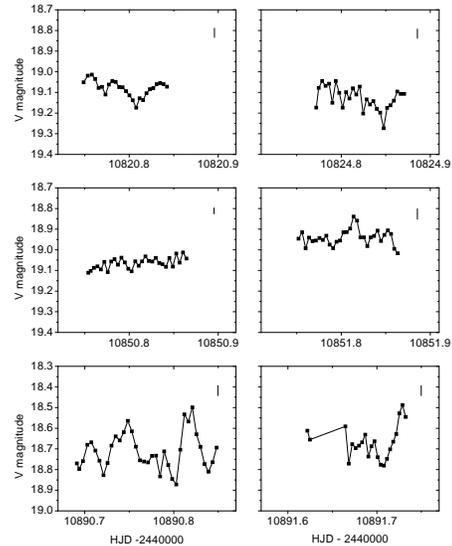,height=3.5in,angle=0}
}}
\caption[]{ 
Differential magnitudes at early 1998. Note the absence of humps in the upper 
panels and 
the different vertical scale in the panels.  A typical error is shown
by  the bar in the upper right corner of every panel.}
\end{figure}

\begin{figure}
\centerline{\hbox{
\psfig{figure=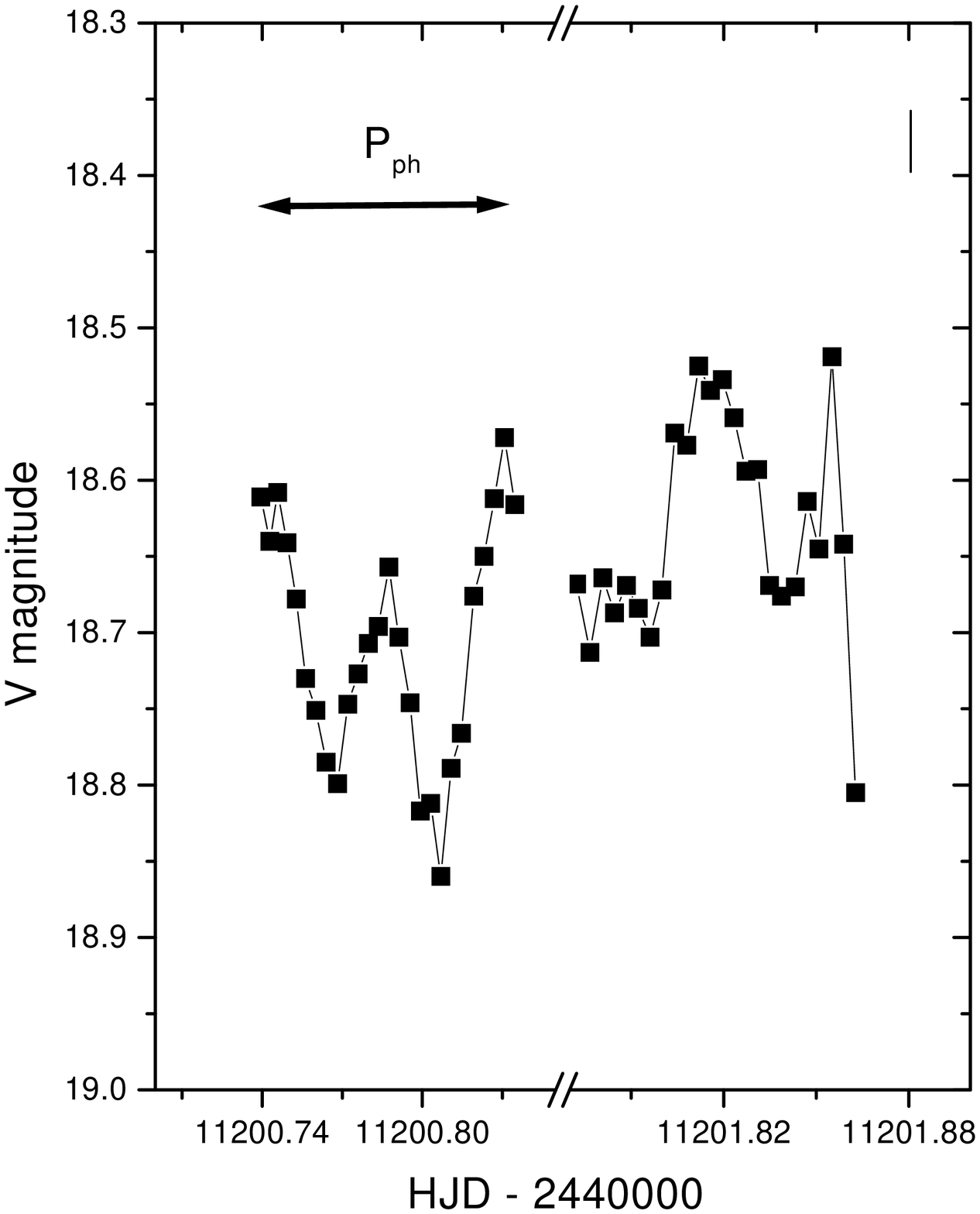,height=3.5in,angle=0}
}}
\caption[]{Differential magnitudes on January 1999. 
The humps appear again. A typical error is shown
by  the bar in the upper right corner. The double arrow shows the length of the 
photometric period.}
\end{figure}

\subsection{The humps: a distinctive character of the light curve}

The differential light curves shown in Fig.\,2 indicate the presence of 
prominent humps on March 1991 and 1995. The humps are roughly symmetrical 
lasting by about 65 minutes and followed by a slow magnitude decrease (March 1991) 
or by 
a secondary low-amplitude hump (March 1995). This picture sharply contrasts
with that observed on early 1998 (Fig.\,3). The humps are completely absent on
January-February 1998 and re-appear (with secondary humps) on March 1998 and 
January 1999 (Fig.\ 4).
A remarkable feature is the absorption like feature seen on January 1998. We will 
show
in the next section that this feature is the ``embrion" of the fully developed 
humps seen two months later.

\subsection{The long-term light curve}

Fig.\,5 shows the long-term light curve of \object{RZ Leo}
during 1987--1999. 
It is evident that the quiescence mean magnitude changes by several tenths of 
magnitude 
in a few years and at 3.5 $\times$ 10$^{-3}$ mag d$^{-1}$ during January and March 
1998.
Unfortunately, the faintness of the object has prevented a continuous
monitoring, so the long-term data are inevitably undersampled.

\begin{figure}
\centerline{\hbox{
\psfig{figure=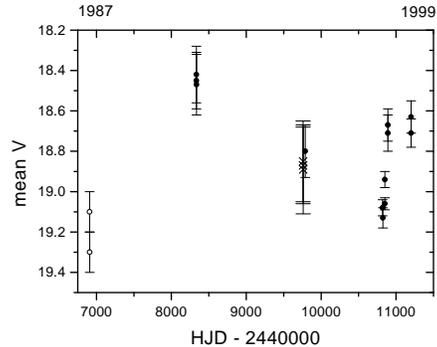,height=3.5in,angle=0}
}}
\caption[]{
Mean magnitudes of \object{RZ Leo} during 1987--1999.  Error bars indicate nightly 
$rms$. Data are from
Howell \& Szkody (1988, open circles), Mennickent (in preparation, crosses) and 
this paper (solid circles).}
\end{figure}

\subsection{Searching for a photometric period}

We removed the long-term fluctuations normalizing the magnitudes to a common 
nightly mean.
Then we applied the Scargle (1982) algorithm, implemented in the $MIDAS$ $TSA$ 
package,
which obeys an exponential probability distribution and is especially useful for
smooth oscillations. In this statistics, the false alarm probability $p_{0}$ 
depends on the periodogram's 
power level $z_{0}$ through $z_{0} \approx \ln{N/p_{0}}$, for small $p_{0}$, where 
$N$ is the number 
of frequencies searched for the maximum power (Scargle 1982, Eq.\,19). 
In our search we used $N$ = 20000, so the  99\% confidence level
(i.e. those corresponding to $p_{0}$ = 0.01) corresponds to a power $z_{0}$ = 
14.5.
The range of frequency scanned was between the Nyquist frequency, i.e.\ 1.7 
$\times 10^{-3}$ c/d and
1 c/d. After applying the method to the whole dataset 
many significant aliases appeared around a period 0\fd076.
Apparently, the light curve was characterized by a
non-coherent or non-periodic oscillation. We decided to start
with our more restricted dataset of March 1991. 
The corresponding periodogram, shown in Fig.\,6, shows a  strong period at 
0\fd0756(12) (108.9 $\pm$ 1.7 m, the error correspond to the half width at 
half maximum of the periodogram's peak) flanked by the $\pm$ 1 c d$^{-1}$ aliases 
at 
0\fd070 (the period found by Howell \& Szkody 1988) and 0\fd082.
The ephemeris for the time of hump maximum is:

\begin{equation}
T_{max} = 244 8333.6186 (35) + 0\fd0756(12) E 
\end{equation}

In order to search for possible period changes we constructed a $O-C$ diagram 
based on timings
obtained measuring the hump maxima. These timings, given in Table 2, were
compared with a test period of 0\fd0756. The $O-C$ differences versus the cycle 
number are 
shown in Fig.\,7. Apparently, the period is not changing
in a smooth and predictable way. In principle, the $O-C$ differences are 
compatible with non-coherent humps and/or 
period jumps.
To explore both possibilities, we searched for seasonal periods. Only datasets of 
March 1991, 1995 and 1998
were dense enough to construct periodograms. The results, given in Table 3,
suggest a non-coherent signal rather than a variable period. In summary, the data 
are compatible with humps repeating with a period of 0\fd0756(12) but in a 
non-coherent way. 
Armed with a photometric period, we constructed seasonal mean light curves.
Only nights with fully developed humps were included.
The results, shown in Fig.\,8, clearly show secondary humps around photometric 
phase 0.5. 
These mean light curves are provided as a hint for future light curve
modeling. 

\begin{table}
\caption[]{Times of hump's maximum. HJD' means HJD - 244\,0000. 
Measures are from our light curves shown in Figs.\,2 to 4 except those
indicated by $^{\dag}$ (derived from the Howell \& Szkody's 1988 light curve) 
and $^{\dag\dag}$ (Mennickent, in preparation). }
\begin{center}
\begin{tabular}{|l|r|} \hline
\multicolumn{1}{c}{HJD'(maximum)} &
\multicolumn{1}{c}{HJD'(maximum)}\\
 \hline \hline
6909.7374(3)$^{\dag}$ &9787.8532(3) \\
6909.8091(3)$^{\dag}$  &10851.8139(3) \\
8333.6186(3) &10890.7484(3) \\
8333.6969(3) &10890.8210(3)\\
8333.7712(3) &10891.7288(6) \\
8334.6789(3) &11200.7458(6) \\
9758.8838(3)$^{\dag\dag}$&11200.8308(6)  \\
9758.8107(3)$^{\dag\dag}$ &11201.8118(3) \\
9787.7798(3) &\\
\hline \hline
\end{tabular}
\end{center}
\end{table}

 \begin{table}
\caption[]{The hump period at different epochs. Our results are 
consistent with a single period.}
\begin{center}
\begin{tabular}{|c|c|} \hline
\multicolumn{1}{c}{Date} &
\multicolumn{1}{c}{period (d)} \\
\hline \hline
March 1991 &0.0756(012) \\
March 1995 &0.0752(150) \\
March 1998 &0.0755(015) \\
\hline \hline
\end{tabular}
\end{center}
\end{table}

\begin{figure}
\centerline{\hbox{
\psfig{figure=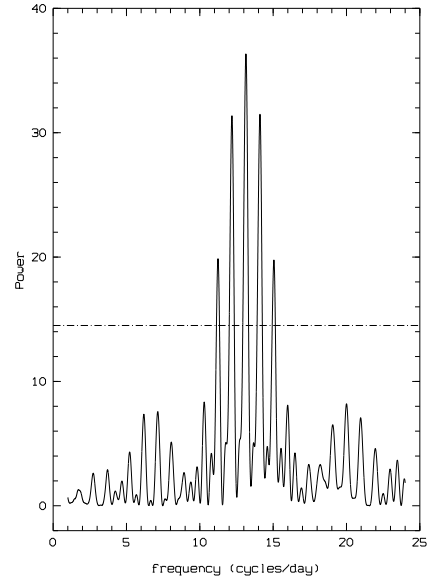,height=3.5in,angle=0}
}}
\caption[]{The Scargle periodogram for the magnitudes of March 1991.
The central peak  at 0\fd0756(12) is flanked by the $\pm$ 1 c\,d$^{-1}$ aliases at
0\fd071 and 0\fd082. The dot-dashed line indicates the 99\% confidence level.}
\end{figure}

\begin{figure}
\centerline{\hbox{
\psfig{figure=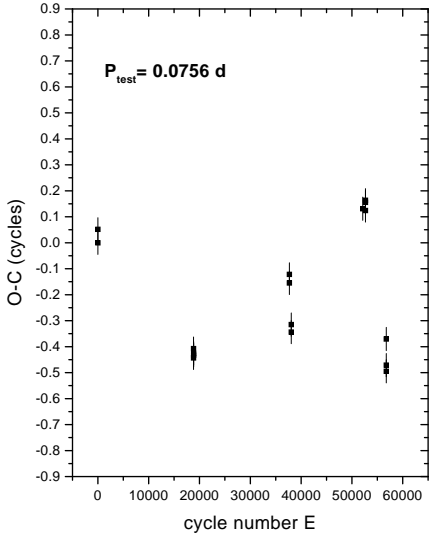,height=3.5in,angle=0}
}}
\caption[]{O - C diagram for the times of hump maximum, 
with respect to a test period of 0\fd0756. Data are from Table 2.
observed in 
ephemeris.
In principle, the figure is compatible with a non-coherent signal or with period 
jumps at some epochs.}
\end{figure}

\begin{figure}
\centerline{\hbox{
\psfig{figure=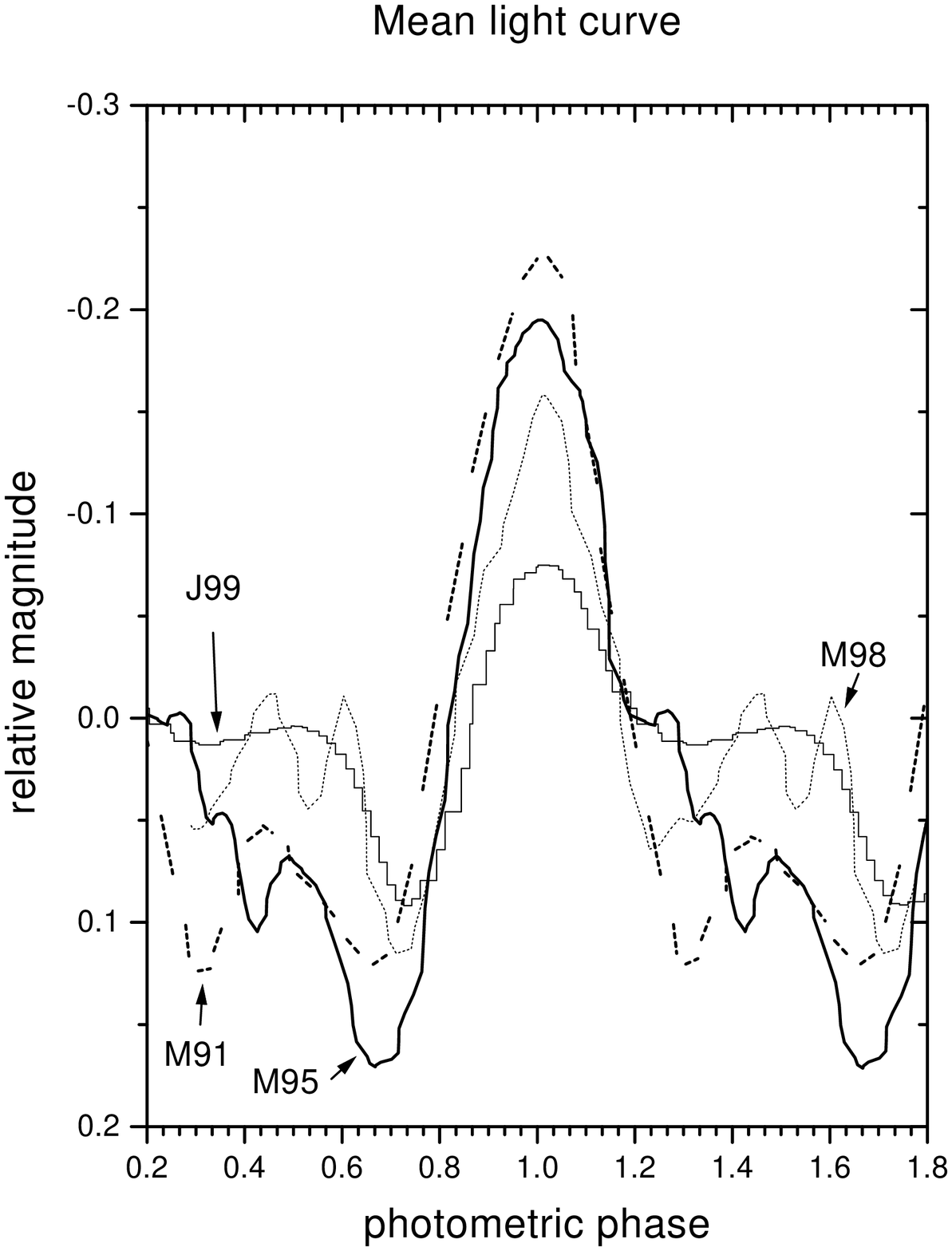,height=3.5in,angle=0}
}}
\caption[]{Mean light curves of fully developed humps, shifted to a common phase 
of maximum. 
The curves are normalized to zero mean intensity. }
\end{figure}
 
\subsection{``Anti-humps" and long-term hump evolution}

A review of the
observations of early 1998 including the diskovery of ``anti-humps"
was given by Mennickent \& Sterken (1999). 
Here we present a more complete analysis of the phenomenon. 
Fig.\ 9 shows in detail the events of early 1998. 
The light curves have been binned with a period 0\fd0756, 
accordingly to Table 3.
The evolution of the hump is singular. It starts as a 0\fm15 absorption feature
(07/01/98) then disappear from the light curve (11/01/98 and 06/02/98) 
and then re-appears like a small wave (07/02/98) and fully developed symmetrical 
hump 
(18/03/98 and 19/03/98). Secondary humps are also visible, with amplitude
roughly 60\% the main hump amplitude. On February 7 a secondary
absorption hump is also visible, along with the main absorption feature.
These ``anti-humps" appear at the same phases where normal humps
develop a month later. A close inspection to the data of February 7
reveals another alternative interpretation:
the observed minima could define the base of the humps.
We have rejected this hypothesis for three reasons:
(1) it does not  fit the ephemeris, indicating a possible shift of the hump 
maximum
by about 0.2 cycles, (2) the peak-to-peak distance between
main and secondary maxima should be 0.3 cycles instead 0.5 cycles, which
is observed the other 3 nights and (3) the secondary maximum should be about 80\%
of the main peak, contrasting with a value of 60\% 
observed other nights. We provide an interpretation for this phenomenon in 
the next Section.

Fig.\,10 shows the hump's amplitude roughly anticorrelated with the
nightly mean magnitude, as occurs in VW Hyi (Warner 1975).  As shown in Fig.\,9,
this anti-correlation is not only due to the increase of hump brightness, but is 
also a true rise of the 
total systemic luminosity, through the whole orbital cycle.
The outlier in Fig.\,10 is a measure by  Howell \& Szkody (1988)
which is a rather doubtful point.
In fact, accordingly to these authors, since their primary goal was to obtain 
differential photometry --
not absolute photometry -- they calibrated their magnitudes using only a few 
standards per night.
They give a formal error of 0\fm03 for the zero point of \object{RZ Leo}, but 
with so few  standards observed, not in the same CCD field,  
it is difficult to control systematic errors
due to variable seeing and atmospheric transparency. In the following,
we will omit this outlier from our diskussion.  
Returning to Fig.\,10, we observe that the  hump disappears
when $V \approx 19$ and attains maximum amplitude when $V \approx 18.4$.
Surprisingly, the hump becomes ``negative" (i.e.\ an absorption feature) when 
the system drops below $\approx$ 19 mag. 
A linear least squares fit to the hump amplitude $\Sigma$ yields:

\begin{equation}
\Sigma = 0.88(8) - 0.82(11) (V-18)
\end{equation}

\noindent where $V$ refers to the nightly mean $V$ magnitude. 

\begin{figure}
\centerline{\hbox{
\psfig{figure=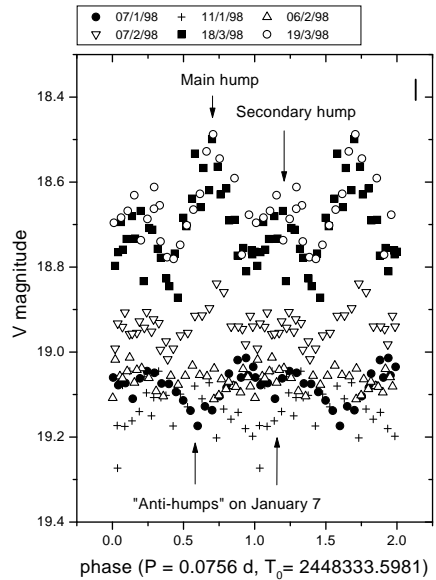,height=3.5in,angle=0}
}}
\caption[]{Light curves of RZ Leo during early 1998 folded with a period 0\fd0756. 
}
\end{figure}

\begin{figure}
\centerline{\hbox{
\psfig{figure=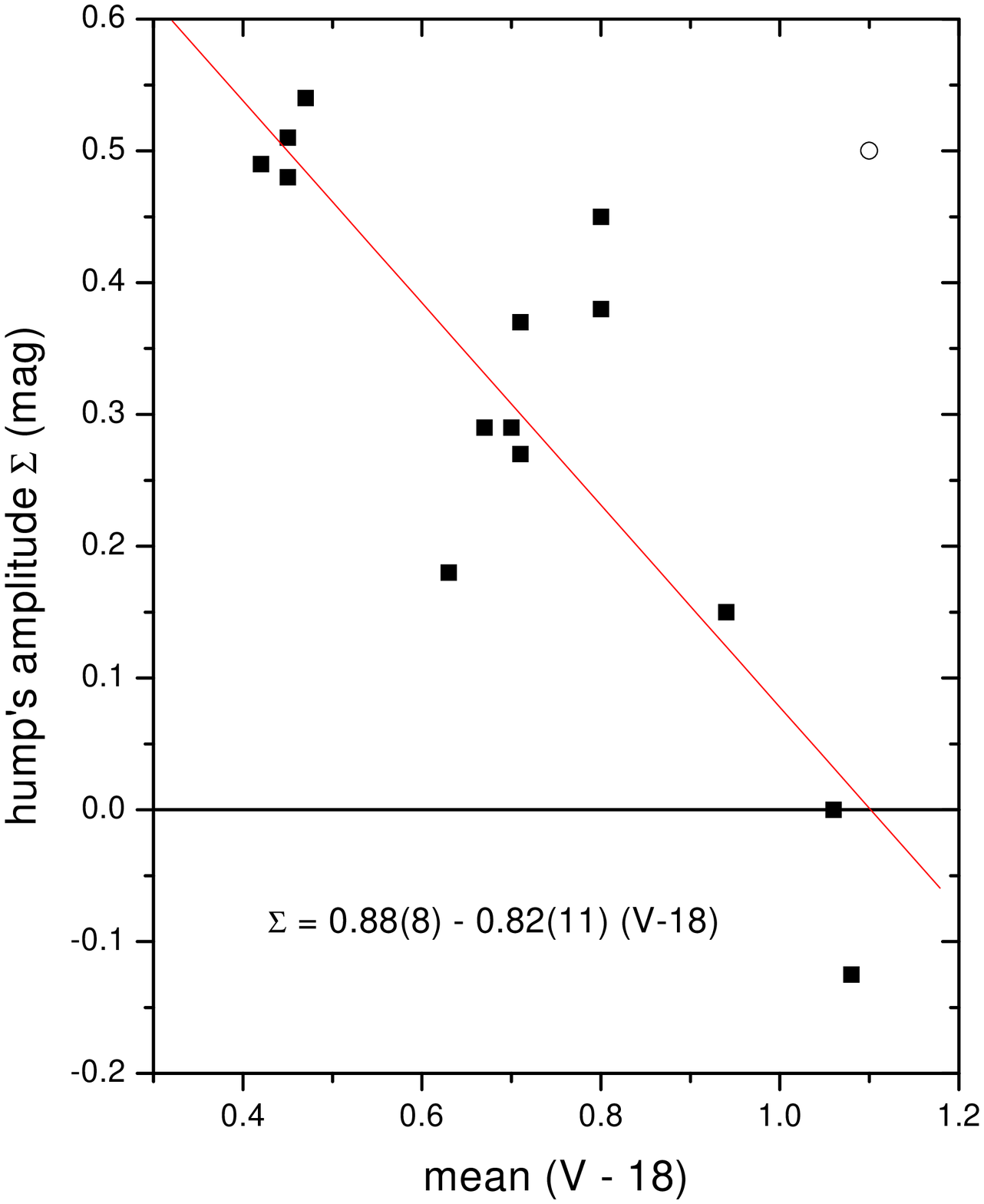,height=3.5in,angle=0}
}}
\caption[]{Hump amplitude versus nightly mean magnitude and the best least-squares 
linear fit.
The outlier by Howell \& Szkody (1988, open circle) is diskussed in the text.}
\end{figure}

\section{Discussion}

\subsection{A Moving hot spot ?}


Any reasonable model for the photometric variability of \object{RZ Leo}
should reproduce the non-coherent humps and their amplitude variations.

It is currently assumed that the humps reflect the release of gravitational energy
when the gas stream hits the accretion disk. The disk's 
luminosity is produced by the same process when disk gas slowly spirals towards
the central white dwarf (e.g.\ Warner 1995a). 

An explanation for the varying humps 
could be a hotpsot moving along the outer disk rim. A bright spot co-rotating with 
the
binary should reflect the binary orbital period, but random 
translations of the hot spot along the outer disk rim should produce 
a non-coherent signal.
Support for this view arises from  the evidence of moving hot spots in some
dwarf novae, e.g.\ \object{KT Per} (Ratering et al.\ 1993) and \object{WZ Sge} 
(Neustroev 1998).

The large scatter observed in the $O-C$ diagram of \object{RZ Leo} (up to 0.4 
cycles) is atypical 
for dwarf novae. For example, \object{U Gem} (Eason et al.\ 1983), \object{IP Peg} 
(Wolf et al.\ 1993)
and \object{V\,2051\,Oph} (Echevarr\'{\i}a \& Alvarez 1993) show quasi-cyclic
period variations, of small amplitude, on time scale of years. In these cases, 
the $O-C$ residuals 
are always lower than 0.02 cycles. The interpretation 
of the changes observed in the above stars is still controversial.

\subsection{\object{RZ Leonis } in the context of WZ Sge stars}

\subsubsection{Evidence for a normal $\alpha$ disk}

In this Section we analyze the events of early 1998.
The observed correlation between the hump amplitude and mean brightness 
might provide important clues on the numerical value of the disk viscosity.

The hot spot and disk bolometric luminosity can be approximated by (Warner 1995a, 
Eq.\ 2.21a and 2.22a):

\begin{equation}
L_{s} \approx \frac{GM_{1}\dot{M}_{2}}{r_{d}}
\end{equation}

\begin{equation}
L_{d} \approx 1/2 \frac{GM_{1}\dot{M}_{d}}{R_{1}}
\end{equation}

\noindent 
where $M_{1}$ and $R_{1}$ are the mass and radius of the primary, 
$r_{d}$ the disk's radius and $\dot{M}_{2}$ and $\dot{M}_{d}$ the
mass {\it transfer} and mass {\it accretion} rates, respectively.

In the following we assume that $L_{s}$ is proportional to 
the hump's peak luminosity and $L_{d}$ is proportional to  the cycle mean 
luminosity.
The disk luminosity so defined includes some contribution from the hot spot,
but it is difficult to exclude
the wide and long-lasting humps from the analysis.
It is apparent from Fig.\,9 and 10 that the increase of hot spot luminosity
is followed by an increase of the disk's luminosity. This effect seems to be true, 
and not simply a
consequence of the hump rising.

Accordingly to Eq.\ 3 and 4, the events of early 1998 may be interpreted as 
follows:
a mass transfer burst starts at the secondary in January 1998 
and then continues with increasing $\dot{M}_{2}$, until March 1998. 
The burst, evidenced in the rising of hump's luminosity in Fig.\,9
triggers  an increase of mass accretion rate inside the disk, as observed in the
rising of the total systemic luminosity. 

The time scale for matter diffusion across the accretion disk, is called 
the viscous time scale (Pringle 1981):

\begin{equation}
t_{\nu} \sim \frac{r_{d}^{2}}{\nu_{K}}
\end{equation}

\noindent
where the viscosity  is given by the Shakura \& Sunyaev (1973) {\it ansatz}:

\begin{equation}
\nu_{K} = \alpha c_{s}H
\end{equation}

\noindent
with $H$ the half-thickness of the disk and $c_{s}$ the sound velocity.
Replacing Eq.\ 6 in 5 and using typical parameters $H/r$ = 0.01, $r = 10^{10}$ cm, 
$c_{s} = 20 \times 10^{5}$ cm s$^{-1}$
we obtain

\begin{equation}
\alpha \sim \frac{5 \times 10^{5}}{t_{\nu}}
\end{equation}

For the diffusion process observed in \object{RZ Leo}
$t_{\nu}$ $\sim$ 6.0 $\times$ 10$^{6}$ s (70 days),
we find $\alpha$ = 0.08, a common value among dwarf novae (Verbunt 1982). 
This value contrasts with the  low $\alpha$ ($<<$ 0.01) invoked to explain the 
long recurrence times
and large amplitude outbursts of some dwarf novae, in particular \object{WZ Sge} 
(Meyer-Hofmeister et al.\ 1998). 
Since our observations indicate a rather normal $\alpha$, the long recurrence time 
must be
explained by another cause. In this context it is worthy to mention the 
hypothesis of inner disk depletion. 

The removal of the inner disk by 
the influence of a magnetosphere (Livio \& Pringle 1992) or the effect of
mass flow via a vertically extended  hot corona above the cool disk 
(also referred as ``coronal evaporation", Meyer \& Meyer-Hofmeister 1994, 
Liu et al.\ 1997, Mineshige et al.\ 1998) 
naturally explains  the long recurrence times. 
Spectroscopic evidence indicates that the inner disk depletion might be a common 
phenomenon 
in SU UMa stars (Mennickent \& Arenas 1998, Mennickent 1999).

\subsubsection{Evidence for a main sequence like secondary}

It has been suggested that many large amplitude dwarf novae have bounced off 
from the orbital period minimum (at $\sim$ 80 min) and are evolving to 
longer orbital periods  with very old, brown-dwarf like secondaries (Howell et 
al.\ 1997). 
This view is supported by the finding of undermassive 
secondaries in  \object{WZ Sge} (Ciardi et al.\ 1998) and V\,592 Her (van 
Teeseling et al.\ 1999) 
and the suspection -- based on the ``superhump" mass ratio -- of this kind
of objects in \object{AL Com} and \object{EG Cnc} (Patterson 1998). In principle,
the large amplitude
and long cycle length of \object{RZ Leo} suggest that this star is an ideal 
candidate for a post-period minimum
system and therefore, for an undermassive secondary. 
Since superhumps have not been yet detected in this star, the only way to 
investigate
this view is analyzing the flux distribution.
We have compiled data from different sources. They are generally non-simultaneous,
and may contain possibly significant variations in the emission of the CVs.
However, to minimize this effect, we have excluded data taken during outburst,
and we have considered data from as few sources as possible and
as close together in time as possible. 

The flux distributions of \object{RZ Leo} and other dwarf noave 
with recognized brown-dwarf like secondaries (and available photometric data)
are compared in Fig.\,11.  The optical-IR flux of a
steady disk, scaled to fit the UBV data of \object{RZ Leo}, is also 
shown.\footnote{In general, the flux distribution of a CV is dominated by the 
accretion disk in optical
wavelengths and by the secondary star in the infrared; the white dwarf and 
boundary layer
mostly contribute to the EUV and X-ray radiation (e.g.\ Frank et al.\ 1992). 
The optical-IR radiation of a infinitely large, steady,  
optically thick disk can be approximated by a $\lambda^{-7/3}$ law (Lynden-Bell 
1969).}

We find that,
in contrast with that observed in the objects with undermassive secondaries,
the flux distribution of \object{RZ Leo} does not drop in the red wavelengths, 
but rises with respect to the disk's contribution. This is 
expected if the secondary were a main-sequence red dwarf.
In fact, the  $V-K$ color of \object{RZ Leo} (viz.\,3.65, Sproats el at.\ 1996)
is representative of a main sequence M0 star (Bessell \& Brett, 1988).
This is consistent with the finding that most secondaries stars for 
cataclysmic variables with $P_{o}$ $<$ 3 h are close to the solar abundance
main sequence defined by single field stars (Beuermann et al.\ 1998).
The above arguments probably rule out the possibility of an undermassive secondary 
in \object{RZ Leo}. 

Our results indicate that 
large amplitude -- long cycle length -- dwarf novae might not necessarily 
correspond to objects in the same evolutive stage.
We have shown that, in spite of the extreme cycle length and outburst amplitude, 
\object{RZ Leo} can not be properly named a {\it \object{WZ Sge} like star}, 
as suggested in the Ritter \& Kolb catalogue (1998).

\begin{figure}
\centerline{\hbox{
\psfig{figure=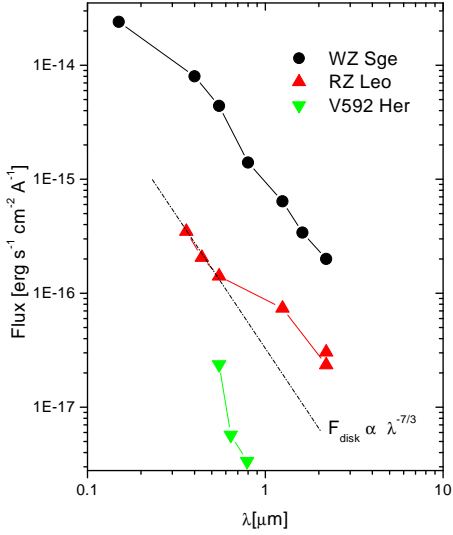,height=3.5in,angle=0}
}}
\caption[]{The flux distribution of three large-amplitude, long cycle-length dwarf 
novae.
Fluxes of \object{WZ Sge} are from Ciardi et al.\ (1998) and references therein. 
Those for 
\object{V592 Her} are based on photometry published by van Teeseling et al.\ 
(1999) and Howell
et al.\ (1991). $UBV$ data for \object{RZ Leo} is from this paper and $JK$ data 
from
Sproats et al.\ (1996) and Szkody (1987). The scaled flux of a steady, optically 
thick, 
accretion disk is given by the dotted line.  }
\end{figure}

\subsection{Anti-humps}

The ratio between hot spot and disk luminosity, for the case of an optically 
thick, 
steady state accretion disk and a simple planar bright spot on the edge 
of the disk, is (Warner 1995a, Eq.\ 2.71):

\begin{equation}
\frac{L^{V}_{s}}{L^{V}_{d}} = f \frac{\tan i}{1+1.5\cos 
i}\frac{\dot{M}_{2}}{\dot{M}_{d}} \frac{R_{1}}{r_{d}} 10^{0.4(B_{sp}-B_{d})}
\end{equation}

\noindent
where $i$ is the systemic inclination, $f$ an efficience factor $\sim$ 1 and 
$B_{sp}$ and $B_{d}$ are the bolometric corrections ($<$ 0) for the spot and disk
respectively. 

Roche-lobe geometry and the assumption of a disk radius 70\% the Roche lobe radius
(an usually good approximation for dwarf novae), yield to:

\begin{equation}
\frac{R_{1}}{r_{d}} \approx \frac{0.40 q^{2/3}}{P_{o}^{3/2}(hr)}
\end{equation}

In addition, hot spot and disk temperatures inferred for dwarf novae indicate
$B_{sp}$ $\sim$ $B_{d}$  (Warner 1995a's diskussion after Eq.\ 2.73 and references 
therein).
Therefore we obtain:

\begin{equation}
\frac{L^{V}_{s}}{L^{V}_{d}} \approx h(i,q,P_{o})  \frac{\dot{M_{2}}}{\dot{M_{d}}} 
\end{equation}

\noindent
where $h (i,q,P_{o}) = \frac{f \tan i}{1+1.5\cos i} \frac{0.40 
q^{2/3}}{P_{o}^{3/2}(hr)}$ is a
function with a numerical value in the range 0.03--0.3 for  most practical 
purposes.
The condition of ``anti-humps" is given by:

\begin{equation}
\frac{L^{V}_{s}}{L^{V}_{d}} < 1
\end{equation}

The above equations suggest that the apparition of ``anti-humps" in a single 
system
depends on the relative values of $\dot{M}_{d}$ and $\dot{M}_{2}$. 
In particular, for \object{RZ Leo}, assuming the orbital and photometric periods 
equals,
a mass ratio of 0.15, i.e.\ representative
for dwarf novae below the period gap (e.g.\ Mennickent et al.\ 1999), and 
a moderate inclination angle of 65$^{o}$, 
this occurs when $\dot{M}_{2} < 6.25 \dot{M}_{d}$ (assuming $f$ =1). The rarity of 
the
phenomenon indicates that  $L_{s} < L_{d}$
is a condition rarely fulfilled among dwarf novae and that $\dot{M}_{2}$ is 
probably always larger or equal
than $ \dot{M}_{d}/h$. Systems with large amplitude humps are candidates for 
$\dot{M}_{2}$
$>$ $ \dot{M}_{d}/h$ whereas high inclination systems with no prominent humps 
(e.g.\ WX Cet, Mennickent
1994) are candidates for $\dot{M}_{2}$ $\sim$ $\dot{M}_{d}/h$.  

We can estimate the mass accretion rate from the recurrence time:\\

\begin{equation}
\dot{M}_{d} = \frac{880}{T_{s}(d)} \times 10^{15} g s^{-1} ~~~~~~T_{s} \leq 900~ d
\end{equation}

\noindent
(Eq.\ 37 by Warner 1995b). Using a supercycle length $T_{s} >$ 2 yr
we obtain $\dot{M}_{d} < 1.2 \times 10^{15}$ g s$^{-1}$. This implies that
$\dot{M}_{2} < 7.5 \times 10^{15}$ g s$^{-1}$ is required to
develop ``anti-humps". This condition is easily satisfied if the mass transfer 
rate
is driven by gravitational radiation, as expected for a dwarf novae below the 
period
gap. In this case, using the system parameters given above, we estimate:\\

\begin{equation}
\dot{M}_{2}^{GR}  = 2.2 \times 10^{15} g s^{-1}
\end{equation} 

\noindent
from Eq.\ 9.20 by Warner (1995a).

Since the mass accretion
rate $\dot{M}_{d}$ is proportional to the viscosity (e.g.\ Cannizzo et al.\ 1998), 
an extremely 
low $\alpha$ disk is not a good site
for developing ``anti-humps". The reason is that, in this case, 
the condition imposed on the mass transfer rate to satisfy Eq.\ 11 is too strong, 
requiring, probably, unrealistic low $\dot{M}_{2}$ values. 
Therefore the presence of ``anti-humps"
in \object{RZ Leo} is consistent with the normal 
$\alpha$ found in the previous section.

\section{Conclusions}

\begin{itemize}
\item The light curve of RZ Leo during a time interval of 11-years is 
characterized by highly variable humps.
\item A non-coherent photometric period of 0\fd0756(2) is consistent with the 
data.
\item The hump amplitude is anti-correlated with the stellar mean brightness.
\item A new phenomenon was reported: the presence of ``anti-humps" when the system 
is faint.
\item Anti-humps might result from a regime of very low mass transfer rate and 
normal alpha disks.
\item Non-coherent humps are compatible with a non-steady hot spot.
\item The rapid response of the accretion disk to the enhanced mass transfer rate 
evidenced by the phenomena
of early 1998 (Fig.\,9) suggests a disk with a normal viscosity parameter $\alpha$ 
$\sim$ 0.08.
\item The possibility of an undermassive secondary is rejected by arguments 
concerning the observed optical and
infrared flux distribution. 
\end{itemize}

\begin{acknowledgements}

This work was partly  supported by Fondecyt 1971064 and DI UdeC  97.11.20-1.
Support for this work was also provided by the National Science Foundation through
grant number GF-1002-98 from the Association of Universities for Research in
Astronomy, Inc., under NSF Cooperative Agreement No. AST-8947990.  C. Sterken 
acknowledges a research grant of the Fund for Scientific Research Flanders (FWO). 
\end{acknowledgements}


\begin{thebibliography}{}
\bibitem[]{} Armitage, P.J., Livio, M., 1998, MNRAS 297, L81
\bibitem[]{} Bessell, M.S., Brett, J.M., 1988, PASP 100, 1134
\bibitem[]{} Beuermann, K., Baraffe, I., Kolb, U., at al.\ 1998, A\&A 339, 518
\bibitem[Bisikalo et al. 1998]{1998ARep...42...33B} Bisikalo, D.V., 
Boyarchuk, A.A., Kuznetsov, O.A., et al.\
1998, Astronomy Reports, 42, 33 
\bibitem[]{} Cannizzo, J.K., Shafter, A.W., Wheeler, J.C., 1988, ApJ 333, 227
\bibitem[]{} Ciardi, D.R., Howell, S.B., Hauschildt, P.H., et al.\ 1998, ApJ 504, 450 
\bibitem[]{} Cristiani, S., Duerbeck, H., Seitter, W.C., 1985, IAU Circ 4027.
\bibitem[Eason et al. 1983]{1983PASP...95...58E} Eason, E. L. E., Africano, J. L., 
Klimke, A., et al.\  1983, PASP 95, 58
\bibitem[Echevarria \& Alvarez 1993]{1993A&A...275..187E} Echevarria, J., Alvarez, 
M., 1993, A\&A 275, 187 
\bibitem[]{} Frank, J. King, A., Raine, D., 1992, Accretion Power in Astrophysics 
(2nd edition), Cambridge University Press
\bibitem[Hessman 1999]{1999ApJ...510..867H} Hessman, F.V., 1999, ApJ 510, 867 
\bibitem[Howell Dobrzycka Szkody \& Kreidl 1991]{1991PASP..103..300H} Howell, S. 
B., Dobrzycka, D. , Szkody, P., et al.\   1991, PASP, 103, 300 
\bibitem[]{} Howell, S.B., 1992, in ``Astronomical CCD Observing and Reduction 
Techniques", ASP conference series vol 23, Ed. S.B. Howell, p. 105
\bibitem[]{} Howell, S.B., Szkody, P., 1988, PASP 100, 224
\bibitem[]{} Howell, S.B., Mitchell, K.J., Warnock III, A., 1988, AJ 95, 247
the Old West: Proceedings
of the 13th North American Workshop on Cataclysmic Variables and Related Objects, 
Eds.\ S. Howell,
E. Kuulkers, C. Woodward, ASP Conference Series, Vol.\ 137, p.\ 9
\bibitem[Howell Rappaport \& Politano 1997]{1997MNRAS.287..929H} Howell, S.B., 
Rappaport, S., Politano, M.,  1997, MNRAS 287, 929 
\bibitem[]{} Livio, M., Pringle, J.E., 1992, MNRAS 259, 23L
198, 383
\bibitem[]{} Lynden-Bell, D., 1969, Nature 223, 690
\bibitem[]{} Mennickent, R.E., 1994, A\&A 285, 979
\bibitem[]{} Mennickent, R.E., 1999 A\&A in press.
\bibitem[]{} Mennickent, R.E., Arenas, J., 1998, PASJ 50, 333
\bibitem[]{} Mennickent, R.E., Sterken, C., 1999, IBVS 4672, 1
\bibitem[Meyer-Hofmeister et al. 1998]{}Meyer-Hofmeister, E., Meyer, F., Liu, B.F., 
1998, A\&A, 339, 507
\bibitem[]{} Meyer, F., Meyer-Hofmeister, E., 1994, A\&A 288, 175
\bibitem[]{} Mineshige, S., Liu, B., Meyer, F., et al.\ 1998, PASJ 50, L5
\bibitem[]{} Misselt, K.A., 1996, PASP 108, 146.
\bibitem[]{} Neustroev, V.V., 1998, Astronomy Reports, 42, 748
\bibitem[]{} Patterson, J., 1998, PASP 110, 1132
\bibitem[]{} Pringle, J.E., 1981, Annual Review of Astronomy and Astrophysics 19, 
137
\bibitem[Ratering Bruch \& Diaz 1993]{1993A&A...268..694R} Ratering, C., Bruch, 
A., Diaz, M., 1993, A\&A 268, 694 
\bibitem[]{} Ritter, H., Kolb, U., 1998, A\&AS 129, 83
\bibitem[]{} Shakura, N.I, Sunyaev, R.A., 1973, A\&A 24, 337
\bibitem[1982]{sca} Scargle, J.D., 1982, ApJ 263, 835
\bibitem[Sproats Howell \& Mason 1996]{1996MNRAS.282.1211S} Sproats, L.N., Howell, 
S B., Mason, K.O., 1996, MNRAS 282, 1211 
\bibitem[Szkody 1992]{1992cvs..work...42S} Szkody, P. 1992, in ``The  Vi\~{n}a del
Mar Workshop on Cataclysmic Variable Stars", Ed. Nikolaus Vogt, ASP Conference Series, Vol.\ 29, p.\ 42
\bibitem[]{} Szkody, P., Howell, S.B., 1991, ApJS 78, 537
\bibitem[]{} Szkody, P., 1987, ApJS 63, 685 
\bibitem[]{} Verbunt, F., 1982, Space Science Review, 32, 379
\bibitem[Van Teeseling Hessman \& Romani 1999]{1999A&A...342L..45V} van Teeseling, 
A., Hessman, F. V.,  Romani, R. W., 1999, A\&A 342, L45 
\bibitem[]{} Vanmuster, T., Howell, S.B., 1996, in ``Cataclysmic Variables and 
Related Objects", Astrophysics and Space Science Library Vol.\,208, Kuwler 
Academic Publishers, Eds. A. Evans and J.H. Wood, p.\,63
\bibitem[]{} Vogt, N., Bateson, F.M., 1982, A\&AS 48, 383
\bibitem[]{} Warner, B., 1975, MNRAS 170, 219
\bibitem[]{} Warner B., 1995a, Cataclysmic Variable Stars, Cambridge University 
Press
\bibitem[]{} Warner, B., 1995b, A\&SS 226, 187
\bibitem[]{} Wolf, S., Mantel, K.H., Horne, K., et al.\ 
1993, A\&A 273, 160 
\bibitem[]{} Wolf, S., Barwig, H., Bobinger, A., et al.\ 1998, 
A\&A 332, 984 
\end{thebibliography}
\end{document}